\documentclass[aps,prl,amsmath,preprintnumbers,twocolumn]{revtex4}

\usepackage{epsfig}
\usepackage{graphics}
\usepackage{latexsym}
\usepackage{amsmath}
\usepackage{amssymb}
\usepackage{rotating}
\usepackage{subfigure}
\usepackage{bm}
\usepackage{color}
\usepackage{amsfonts}
\usepackage{amsmath}
\usepackage{mathtools}

\begin{document}

\preprint{ADP-15-16/T918}

\title{Pure sea-quark contributions to the magnetic form factors of $\Sigma$ baryons}

\author{P. Wang$^{1,2}$}
\author{D. B. Leinweber$^{3}$}
\author{A. W. Thomas$^{3,4}$}

\affiliation{$^1$Institute of High Energy Physics, CAS,
		 Beijing 100049, China}

\affiliation{$^2$Theoretical Physics Center for Science Facilities, CAS,
		 Beijing 100049, China}

\affiliation{ $^3$Special Research Center for the Subatomic Structure
  of Matter (CSSM), Department of Physics, University of
  Adelaide, SA 5005, Australia}

\affiliation{ $^4$ ARC Centre of Excellence in Particle Physics
at the Terascale,
Department of Physics, University of Adelaide, SA 5005, Australia}

\begin{abstract}

We propose the pure sea-quark contributions to the magnetic form
factors of $\Sigma$ baryons, $G_{\Sigma^-}^u$ and $G_{\Sigma^+}^d$, as
priority observables for the examination of sea-quark contributions to
baryon structure, both in present lattice QCD simulations and possible
future experimental measurement.  $G_{\Sigma^-}^u$, the $u$-quark
contribution to the magnetic form factor of $\Sigma^-$, and
$G_{\Sigma^+}^d$, the $d$-quark contribution to the magnetic form
factor of $\Sigma^+$, are similar to the strange quark contribution to
the magnetic form factor of the nucleon, but promise to be larger by
an order of magnitude.  We explore the size of this quantity within
chiral effective field theory, including both octet and decuplet
intermediate states.  The finite range regularization approach is
applied to deal with ultraviolet divergences. Drawing on an
established connection between quenched and full QCD, this approach
makes it possible to predict the sea quark contribution to the
magnetic form factor purely from the meson loop.  In the familiar
convention where the quark charge is set to unity $G_{\Sigma^-}^u =
G_{\Sigma^+}^d$.  We find a value of $-0.38^{+0.16}_{-0.17}\ \mu_N$,
which is about seven times larger than the strange magnetic moment of the
nucleon found in the same approach.  Including quark charge factors,
the $u$-quark contribution to the $\Sigma^-$ magnetic moment exceeds
the strange quark contribution to the nucleon magnetic moment by a
factor of 14.

\end{abstract}

\maketitle


It is well known that a complete characterization of baryon
substructure must go beyond three valence quarks.  Strange quark
contributions to the properties of the nucleon have attracted a lot of
interest since the originally puzzling EMC results concerning the
proton spin~\cite{EMC}. While that particular motivation has
faded~\cite{Myhrer:2007cf,Thomas:2008bd,Thomas:2008ga}, the role of
the sea remains a central issue in QCD, especially with respect to
lattice QCD. There such terms involve so-called ``disconnected
graphs''; that is, quark loops which are connected only by gluons to
the valence quarks. Despite enormous effort~\cite{Lewis:2002ix}, only
one direct lattice QCD calculation has produced a non-zero
result~\cite{Doi:2009sq}, albeit with errors which mean that it is not
statistically different from zero.

With very few exceptions, the form factor studies, which complement
the recent experimental progress at facilities such as Jefferson Lab,
deal with so called connected contributions, in which the external
current acts on a quark line running directly from the hadronic source
to sink.  As discussed below, only a few studies have directly
addressed the disconnected contributions.  Perhaps the most famous
example of a disconnected contribution is the strange quark
contribution to the nucleon elastic form
factors~\cite{Leinweber:1995ie,Leinweber:1999nf,Leinweber:2004tc,%
Leinweber:2005bz,Leinweber:2006ug,Wang:2009ta,Thomas:2012tg,Wang:2014nhf,Shanahan:2014tja}.
Its fundamental importance is associated with the fact that it is
directly analogous to the vacuum polarization contribution to the Lamb
shift, the correct calculation of which confirmed the validity of
Quantum Electrodynamics.

Parity-violating electron scattering (PVES) has proven to be a
valuable tool for experimentally determining the strange quark
contribution to the electromagnetic form factors of the proton.  Under
the assumption of charge symmetry, one can deduce the strange electric
or magnetic form factor ($G_{E,M}^s(Q^2)$ ) from measurements of the
corresponding proton and neutron electromagnetic form factors {\it
  and} the neutral-weak vector form factor of the proton, through its
contribution to PVES. While PVES measurements are very challenging, a
number of groups have succeeded, starting with SAMPLE at
Bates~\cite{Spayde:2003nr} and then A4 at Mainz~\cite{a4,Maas:2004dh}
and G0~\cite{Armstrong:2005hs} and
HAPPEX~\cite{Acha:2006my,Aniol:2005zg,Aniol:2004hp} at Jefferson
Lab. Up to now, the experiments have not provided an unambiguous
confirmed answer to the sign of the strange form factors, although
global analyses do tend to suggest that $G_M^s(0) < 0$ is
favoured~\cite{Young:2006jc,Gonzalez-Jimenez:2014bia}.

Though there exist predictions of the strange form factors from
lattice with the help of charge
symmetry~\cite{Leinweber:2004tc,Leinweber:2005bz,Leinweber:2006ug,Wang:2009ta,
Thomas:2012tg,Wang:2014nhf,Shanahan:2014tja},
it is difficult to simulate this quantity directly because it is
purely from the disconnected diagrams and is also quite small.  Even
an unambiguous determination of the sign of the strange form factor is
an important step in the quest to understand the structure of
nucleon. It is related to how the strange and anti-strange quarks are
distributed in the nucleon. The sign of the strange form factor will
shed light on whether the $qqqs\bar{s}$ components in nucleon is
dominated by colored di-quark configurations or by color singlet
configurations~\cite{Zou}.

In this Letter, we propose that the quantity $G_{\Sigma^-}^u$, the $u$-quark
contribution to the magnetic form factor of $\Sigma^-$ or similarly
$G_{\Sigma^+}^d$, the $d$ quark contribution to the magnetic form
factor of $\Sigma^+$, is equally important to $G_M^s(0)$.  Because the
light quark mass of the $u$ or $d$ quark governs the magnitude of the
contribution, it is expected to be larger and less difficult to
measure in lattice QCD.  It is similar to the strange form factor in
the sense that both of the quantities arise purely from
``disconnected'' sea-quark contributions. However, in an effective
field theory framework, $G_{\Sigma^-}^u$ and $G_{\Sigma^+}^d$ are
generated by a $\pi$ meson loop, which should be much larger than the
strange form factor generated from a $K$ meson loop.  They will serve
as an ideal quantity for future lattice simulations and will shed
light on the sea quark properties of baryons.

Chiral effective field theory (EFT) is a useful tool with which to
study hadron properties at low energy. There has been some work on
strange form factors with heavy baryon chiral
EFT~\cite{Hemmert1,Hemmert2}.  However, there is an unknown low energy
constant appearing in the chiral Lagrangian, which has limited the
capacity to calculate the strange magnetic form factor.  In other
words, the quantity one wishes to predict -- the strangeness vector
current matrix element -- is the same quantity one needs to know in
order to make a prediction~\cite{Musolf,Kubis}.  While this is the
case in conventional chiral EFT, experience with
finite-range-regularization (FRR), has shown that by varying the
regulator parameter, one can model the shift in strength from the loop
contributions into the core.  This suggests that within FRR $\chi$-EFT
one might identify the core contribution with the tree level
contribution and make the approximation that, for $\Lambda$ in the
region of 0.8 GeV, the sea quark content of the core is negligible.
In this way, full QCD results have been obtained rather successfully
from quenched lattice
data~\cite{Leinweber:2004tc,Leinweber:2005bz,Leinweber:2006ug,Wang:2009ta,Wang:2008vb,Wang:2012hj}.
We should emphasize that unquenching only works for the particular
choice of regulator mass, $\Lambda$ around 0.8 GeV, because only then
does one define a core contribution that is approximately invariant
between quenched and full QCD.

We will apply heavy baryon chiral effective field theory with finite
range regularization to study the pure sea-quark contribution to the
magnetic form factors of $\Sigma$ baryons.  In presenting the
formalism, we choose to focus on the $d$-quark contribution to
$\Sigma^+$ form factors.  This channel is very similar to the
$s$-quark contribution to the proton.  In the standard convention
where the quark charge is set to unity $G_{\Sigma^-}^u =
G_{\Sigma^+}^d$.

In heavy baryon chiral EFT,
the lowest order chiral Lagrangian for the baryon-meson interaction
which will be used in the calculation of the magnetic form factor,
including the octet and decuplet baryons, is expressed as
\begin{eqnarray}\label{lol}
{\cal L}_v &=&2D{\rm Tr}\bar{B}_v S_v^\mu\{A_\mu,B_v\}
+2F{\rm Tr}\bar{B}_v S_v^\mu[A_\mu,B_v]
\nonumber \\
&& +{\cal C}(\bar{T}_v^\mu A_\mu B_v+\bar{B}_v A_\mu T_v^\mu),
\end{eqnarray}
where $S_\mu$ is the covariant spin-operator defined as
\begin{equation}
S_v^\mu=\frac i2\gamma^5\sigma^{\mu\nu}v_\nu.
\end{equation}
Here, $v^\nu$ is the baryon four velocity (in the rest frame, we have
$v^\nu=(1,0)$) and  
$D$, $F$ and $\cal C$ are the usual SU(3) coupling constants.
The chiral covariant derivative, $D_\mu$, is written as $D_\mu
B_v=\partial_\mu B_v+[V_\mu,B_v]$. The pseudoscalar meson octet
couples to the baryon field through the vector and axial vector
combinations
\begin{equation}
V_\mu=\frac12(\zeta\partial_\mu\zeta^\dag+\zeta^\dag\partial_\mu\zeta),~~~~~~
A_\mu=\frac12(\zeta\partial_\mu\zeta^\dag-\zeta^\dag\partial_\mu\zeta),
\end{equation}
where
\begin{equation}
\zeta=e^{i\phi/f}, ~~~~~~
f=93~{\rm MeV}.
\end{equation}

\begin{figure}[t]
\includegraphics[width=7cm]{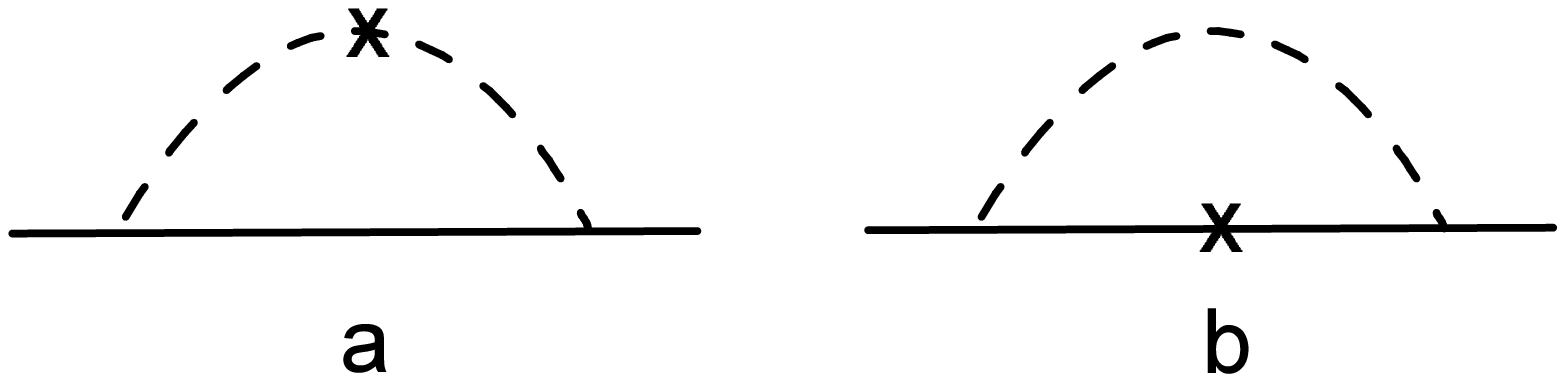}
\caption{Feymann diagrams for the calculation of the magnetic form
  factor of the $\Sigma^+$. Diagrams a and b correspond to the
  leading- and next-to-leading-order diagrams, respectively.}
\label{fig:1}
\end{figure}

As explained above, following earlier successful studies of the
connection between quenched and full QCD, our working hypothesis is
that the $d$ quark contribution to the magnetic form factor of the
$\Sigma^+$ comes purely from the meson loop diagram, which is shown in
Fig.1. There are two types of diagram.  {}Fig.~1a is the leading order
contribution, where the external field couples to the meson. Fig.~1b
is the next-to-leading order contribution, where the external field
couples to the baryon.  That the $K$ meson loop provides a very small
contribution to the magnetic form factor was shown in the previous
study of the strange magnetic form factor~\cite{Wang:2014nhf}.  Here
we consider the $\pi$ loop contribution.  Both octet and decuplet
intermediate states are included.  The contribution from the process
shown in Fig.~1a is expressed as
\begin{equation}
G^{d\,{\rm (1a)}}_{\Sigma^+} = P_{\pi^+\Sigma^0} + P_{\pi^+\Lambda} + P_{\pi^+\Sigma^{*0}},
\end{equation}
where the respective terms correspond to the 
intermediate $\Sigma^0$, $\Lambda$ and $\Sigma^{*0}$ states.
$P_{\pi^+\Sigma^0}$ can be obtained as
\begin{equation}
P_{\pi^+\Sigma^0} = -\frac{m_\Sigma\, F^2}{12\, \pi^3\, f_\pi^2} \int
d^3k \frac{k^2\, u_1 \, u_2}{\omega^2_1\, \omega^2_2}.
\end{equation}
In the now standard notation, $u_1$ ($u_2$) is the regulator
introduced in the finite range regularization with momentum $\vec{k}_1
= \vec{k} + \vec{q}/2$ ($\vec{k}_2= \vec{k} - \vec{q}/2$).  $\omega_1$
($\omega_2$) is the energy of a pion with momentum $\vec{k}_1$
($\vec{k}_2$).  The charge of the $d$ quark has been set to unity,
consistent with the universal convention when discussing the strange
quark form factors of the proton.

The intermediate $\Lambda$ contribution in Fig.~1a has 
the following relationship with the $\Sigma^0$
\begin{equation}
P_{\pi^+\Lambda} = \frac{D^2}{3F^2}\, P_{\pi^+\Sigma^0}.
\end{equation}
For the decuplet part, the contribution is written as
\begin{equation}
P_{\pi^+\Sigma^{*0}} = \frac{m_\Sigma\, {\cal C}^2}{432\, \pi^3\, f_\pi^2} 
\int d^3k \frac{k^2\, u_1\, u_2\, ( 1 +
  \Delta/(\omega_1+\omega_2))}{\omega_1\, \omega_2\, (\omega_1 +
  \Delta)\, (\omega_2 + \Delta)},
\end{equation}
where $\Delta$ is the mass difference between the $\Sigma^{*0}$ and $\Sigma^0$.

The next-to-leading order contribution of Fig.~1b is
\begin{equation}
\label{NLO}
G^{d\,{\rm (1b)}}_{\Sigma^+} = P_{\Sigma^0}\, \mu^d_{\Sigma^0} +
P_{\Sigma^{*0}}\, \mu^d_{\Sigma^{*0}} +
P_{\Sigma^{*0}\Sigma^0(\Lambda)}\, \mu^d\, .
\end{equation}
This includes octet, decuplet and octet-decuplet transition contributions in Fig.~1b.
The octet contribution arising from the $\Sigma^0$ is written as
\begin{equation}
P_{\Sigma^0} = \frac{F^2}{16\, \pi^3\, f_\pi^2} \int d^3k \,
\frac{k^2\, u^2_k}{\omega^3_k} \, ,
\end{equation}
corresponding to the $\Sigma^0$ state appearing in the configuration
$\pi^+\Sigma^0$.  The decuplet contribution from the $\Sigma^{*0}$ is
obtained as
\begin{equation}
P_{\Sigma^{*0}} = - \frac{5\, {\cal C}^2}{864\, \pi^3\, f_\pi^2} \int
d^3 k\, \frac{k^2\, u^2_k}{\omega_k\, (\omega_k+\Delta)^2} \, .
\end{equation}
The ${\Sigma^{*0}\Sigma^0(\Lambda)}$ transition contribution 
to the magnetic form factor is written as
\begin{equation}
P_{\Sigma^{*0}\Sigma^0(\Lambda)} = - \frac{(D-F)\, {\cal C}}{36\,
  \pi^3\, f_\pi^2} \int d^3k\, \frac{k^2\, u^2_k}{\omega^2_k\,
  (\omega_k+\Delta)} \, .
\end{equation}
Here $\mu^d_B$ is the $d$ quark contribution to the magnetic moment of 
the baryon $B$ at tree level, i.e.
\begin{equation}
\mu^d_{\Sigma^0} = \frac23 \, \mu^d_{\Sigma^{*0}} = \frac23 \, \mu_d.
\end{equation}
{}For the last term in Eq.~(\ref{NLO}), the following transition moments is applied
\begin{equation}
\mu^d_{\Sigma^0\Sigma^{*0}} = \frac{\sqrt{3}}{3} \,
\mu^d_{\Lambda\Sigma^{*0}} =
\frac{\sqrt{2}}3\, \mu_d\, .
\end{equation}

\begin{table}[t]
\caption{Pure sea-quark contributions to the magnetic moments of
  $\Sigma$ baryons, $G_{\Sigma^-}^u$ or $G_{\Sigma^+}^d$. Values are
  for unit charge sea quarks in $\mu_N$.  The dependence of the
  results on the finite-range regulator parameter, $\Lambda$ is
  presented.}
\begin{ruledtabular}
\begin{tabular}{ccccccc}
$\Lambda$ (GeV) & 0.6 & 0.7 & 0.8 & 0.9 & 1.0   \\ \hline
LO & $-0.21$ &  $-0.27$ & $-0.34$ & $-0.42$ & $-0.49$ \\
NLO & $-0.017$ &  $-0.025$ & $-0.035$ & $-0.045$ & $-0.057$ \\
$G_{\Sigma^-}^u\,$or $G_{\Sigma^+}^d$ & $-0.22$ &  $-0.30$ & $-0.38$ & $-0.46$ & $-0.55$ \\
\end{tabular}
\end{ruledtabular}
\end{table}

In the numerical calculations, the parameters are chosen as $D=0.76$
and $F=0.50$ ($g_A=D+F=1.26$). The coupling constant ${\cal C}$ is
chosen to be $-2D$. The form of the regulator function, $u(k)$, could
be chosen to be a monopole, dipole or Gaussian function, any of which
would give similar results~\cite{Young:2002ib}.  In our calculations,
a dipole form is chosen because that is the empirical shape of the
nucleon axial form factor~\cite{Guichon:1982zk}
\begin{equation}
u_k=\frac1{(1+k^2/\Lambda^2)^2} \, ,
\end{equation}
with $\Lambda = 0.8\pm 0.2$ GeV. 

As we explained earlier, this choice has been widely applied in the
extrapolation of lattice data for hadron mass, moments, form factors,
radii, first moments of GPDs,
etc.~\cite{Young:2002ib,Leinweber3,Wang:2014nhf,Wang4,Wang5,Allton:2005fb,Armour:2008ke,Hall1,Hall2}.
With this choice it has been shown that reasonable physical results
can be obtained from the quenched lattice data at both leading and
next leading
order~\cite{Young:2002ib,Leinweber:2006ug,Leinweber3,Leinweber:2004tc,Leinweber:2005bz,Wang:2009ta,Wang:2008vb,Wang:2012hj,Wang:2014nhf}.
$\Lambda$ around 0.8 GeV is the value required to identify a core
contribution that is invariant between quenched and full QCD.  This
invariance of the core is based upon the assumption that the 3-quark
core of the $\Sigma^+$ contains no $d$ quark component.

While our calculation is motivated by chiral effective field theory
with the same chiral Lagrangian, our calculation with FRR is at a
physically motivated scale, where earlier work has suggested that the
residual series of analytic terms best describes the three-quark core
contributions. From the previous extrapolation of quenched lattice
data, it is found that this preferred value of $\Lambda$ in the dipole
regulator is around 0.8 GeV. The variation of $\Lambda$ from 0.6 to 1
GeV provides an estimate of the degree of model dependence of our
result.

The contribution of the pure sea-quark contribution to the $\Sigma$
magnetic moment at leading and next-to-leading order is shown in Table
I.  The leading order diagram shown in Fig.\ 1(a) gives a negative
contribution to the magnetic form factor.  The contributions from the
next-to-leading order diagrams are much smaller than the leading
contribution.  They depend on the parameter $\mu_d$.  Assuming SU(3)
symmetry, one has
$\mu_d=\mu_s=-\frac{1}{2}\mu_u=-\frac{1}{3}\mu_D$. In fact, this
relation was applied in our previous investigation of nucleon magnetic
form factors \cite{Wang:2012hj,Wang4}.  In the previous extrapolation
of nucleon magnetic form factors, we found $\mu_D$ equal
$2.55$~$\mu_N$ and $2.34$~$\mu_N$ for full QCD and quenched QCD
extrapolations, respectively~\cite{Wang:2012hj,Wang4}.  Therefore,
$\mu_d=-0.8$~$\mu_N$ should be a good estimate.

\begin{figure}[t]
\includegraphics[width=8.0cm]{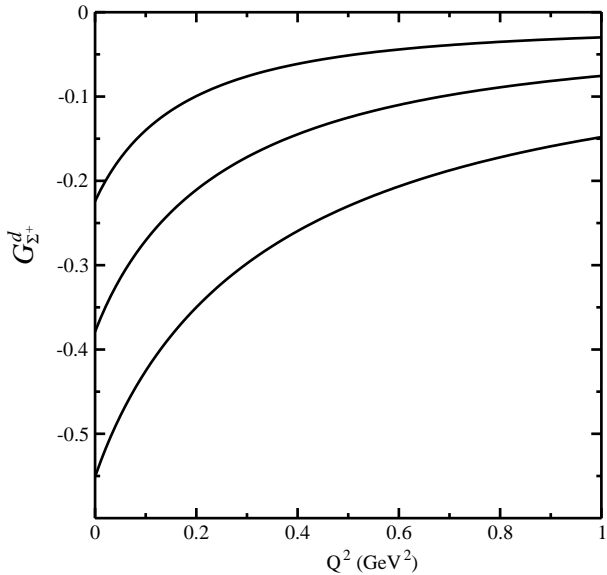}
\caption{The $Q^2$ dependence of the $d$-quark contribution to the
  magnetic form factor of $\Sigma^+$. The upper, middle and lower
  lines are for $\Lambda=0.6$, 0.8 and 1.0 GeV, respectively.  In the
  standard convention $G^d_{\Sigma^+} = G^u_{\Sigma^-}$.}
\label{fig:2}
\end{figure}

In Fig.~2, we show the magnetic form factor $G_{\Sigma^+}^d(Q^2)$
versus $Q^2$ at $\Lambda=$ 0.6, 0.8 and 1.0 GeV.  One can see that
$G_{\Sigma^+}^s(Q^2)$ decreases in magnitude with the increasing
$Q^2$.  It is obvious that the magnetic form factor does not change
sign for any of the choices of $\Lambda$ when $Q^2$ increases.  This
is just like the strange magnetic form factor of the nucleon.
However, the absolute value of $G_{\Sigma^+}^d(Q^2)$ is about one
order of magnitude larger than $G_{N}^s(Q^2)$.  Since its absolute
value decreases with the increasing $Q^2$, it would be preferable to
attempt to measure the magnetic form factor at low $Q^2$. For example,
when $Q^2$ is less than 0.2 GeV$^2$, the absolute central value of
$G_{\Sigma^+}^d$ is larger than 0.2 $\mu_N$ .

At $Q^2=0$, the $d$ quark contribution to the magnetic moment of the
$\Sigma^+$ is $\mu_{\Sigma^+}^d=G_{\Sigma^+}^d(0)=-0.38~\mu_N$.  If we
vary $\Lambda$ from 0.6 GeV to 1 GeV, $\mu_{\Sigma^+}^d$ will change
from $-0.22$~$\mu_N$ to $-0.55$~$\mu_N$. Numerical results show that
$\mu_{\Sigma^+}^d$ remains negative over a large parameter
range. Compared with the strange magnetic moment of the proton, the
value of $\mu_{\Sigma^+}^d$ is about seven times
larger~\cite{Leinweber:2004tc,Wang:2014nhf}.

For unit charge sea-quarks, $G^d_{\Sigma^+} = G^u_{\Sigma^-}$.  Thus
the magnitude of the sea-quark contribution further doubles in an
experimental measurement of the contribution of the $u$ quark to the
form factor of $\Sigma^-$.

Motivated by the importance of establishing the properties of
disconnected contributions to physical quantities in lattice QCD, we
have shown that $G_{\Sigma^-}^u(Q^2)$ and $G_{\Sigma^+}^d(Q^2)$ have
the practical advantage that their values are much larger than the
strange magnetic form factor of the nucleon.  Since the absolute value
of $G_{\Sigma^+}^d(Q^2)$ is nearly one order of magnitude larger than
the strange magnetic form factor of the nucleon, it would clearly be
better to simulate this quantity in place of the strange form factor of
the nucleon.  

Since the lattice simulations will almost certainly be made over a
range of light quark masses, we have investigated the pion mass
dependence of $G_{\Sigma^+}^d(0)$.  The results are shown in Fig.~3,
where the upper, middle and lower lines are for $\Lambda=$ 0.6, 0.8
and 1 GeV, respectively.  From the figure, one can see that with
increasing quark mass the absolute value of $\mu_{\Sigma^+}^d$
decreases.  However, even at $m_\pi^2$ = 0.2 GeV$^2$,
$\mu_{\Sigma^+}^d$ is still much larger than the strange magnetic
moment of the nucleon at the physical pion mass.

An additional feature of the $\Sigma$ baryon is the presence of a
strange quark in the two-point correlation function.  In calculating
the disconnected sea-quark contribution, one multiplies the
disconnected loop by the standard two-point function in creating the
full three-point function.  The presence of a strange quark in the
two-point function will assist in reducing statistical noise in the
three-point correlation function for the pure sea-quark contribution.

\begin{figure}[t]
\includegraphics[width=8.2cm]{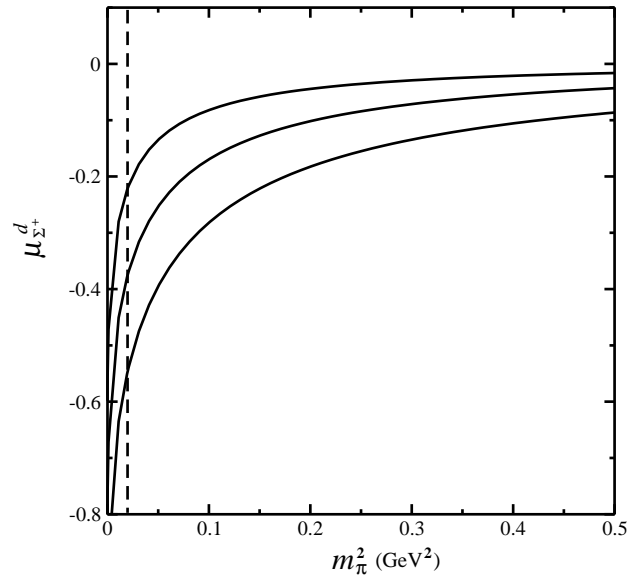}
\caption{The pion mass dependence of the $d$-quark contribution to the
  magnetic moment of the $\Sigma^+$. The upper, middle and lower lines
  are for $\Lambda=0.6$, 0.8 and 1.0 GeV, respectively.  In the
  standard convention $G^d_{\Sigma^+} = G^u_{\Sigma^-}$.}
\label{fig:3}
\end{figure}

Given that $G_{\Sigma^+}^d(Q^2)$ is dominated by the contribution of a
$\pi$ meson loop and having strange quarks in the two-point
correlation function is advantageous, one might also consider the $d$
quark contribution to the magnetic form factor of the $\Xi^{0}$ or the
the $u$ quark contribution to the magnetic form factor of the
$\Xi^{-}$.  These quantities are also determined by a $\pi$ meson
loop.  However, the coupling of $\pi$ and $\Xi^{0}$ is much smaller
resulting in a very small value of $G_{\Xi^0}^d(Q^2)$. Thus
$G_{\Sigma^+}^d$ has unique advantages with respect to studies of the
contributions to the structure of baryons through disconnected sea
quark terms.

In summary, we have argued the importance of studying the pure
sea-quark contributions to $\Sigma$-baryon form factors,
$G_{\Sigma^-}^u(Q^2)$ and $G_{\Sigma^+}^d(Q^2)$.  Because of the
significant enhancement associated with the light $u$ or $d$ quarks,
these observables have distinct quantitative advantages over the
strange form factors of the nucleon.  This enhancement arises because
the pure light sea-quark contribution to the magnetic form factors of
$\Sigma$ baryons is dominated by the $\pi$-meson cloud contribution.
This is much larger than the nucleon strange magnetic form factor
which originates in the $K$-meson cloud.
 
We calculated $G_{\Sigma^-}^u(Q^2)$ and $G_{\Sigma^+}^d(Q^2)$ within
heavy baryon chiral effective field theory including both octet and
decuplet intermediate states.  The pure sea-quark contribution to the
magnetic moment is $G_{\Sigma^-}^u(Q^2) = G_{\Sigma^+}^d(Q^2)=
-0.38^{+0.16}_{-0.17}\ \mu_N$, which is about seven times larger than
the nucleon strange magnetic moment and 14 times larger for
$G_{\Sigma^-}^u(Q^2)$ in experiment.

We also calculated the pion mass dependence of the pure sea-quark
contributions.  When the pion mass is about 300-400 MeV, the absolute
value of $\mu_{\Sigma^+}^d$ is still around $0.2$ $\mu_N$. It seems
likely that future lattice simulations may be able to determine
$G_{\Sigma^+}^d$ directly with more accuracy than the strange form
factor of the nucleon, $G_N^s$.  The value or even the sign of
$G_{\Sigma^+}^d(Q^2)$ would be very helpful in pinning down the size
and origin of five-quark configurations in baryons.

\section*{Acknowledgments}

This work was supported in part by DFG and NSFC (CRC 110), by the
National Natural Science Foundation of China (Grant No. 11475186) and
by the Australian Research Council through grants FL0992247 (AWT),
DP140103067 and DP150103164 (DBL) and through the ARC Centre of
Excellence for Particle Physics at the Terascale.

\end{document}